\documentstyle[aps,preprint,prb]{revtex}
\begin{document}

\title{An Anderson Impurity in a Semiconductor} 
\author {Clare C. Yu and M.\ Guerrero$^{\dagger}$}
\address{
Department of Physics and Astronomy, 
University of California, Irvine, CA 92717-4575} 
\date{ \today}

\maketitle
\setcounter{page}{0}
\thispagestyle{empty}
\begin{abstract}
We study an Anderson impurity in a semiconducting host using 
the density
matrix renormalization group technique. We use the $U=0$ 
one--dimensional Anderson Hamiltonian at half 
filling as the semiconducting host
since it has a hybridization gap. By varying the
hybridization of the host, we can control 
the size of the semiconducting gap. We consider chains
with 25 sites and we place the Anderson impurity (with $U>0$)
in the middle of the chain. We dope the half--filled system with 
one hole
and we find two regimes: when the hybridization of the impurity is 
small, the hole density
and the spin are localized near the impurity. When the 
hybridization of the impurity is large, 
the hole and spin density are spread over the lattice.
Additional holes avoid the impurity and are extended throughout
the lattice. Away from half--filling, the semiconductor with 
an impurity is analogous to
a double well potential with a very high barrier.
We also examine the chemical potential as a function of 
electron filling,
and we find that the impurity introduces midgap states when the
impurity hybridization is small.
\end{abstract}

\vspace{1.0cm}
$^{\dagger}$ Current address: T-11, MS B262, Los Alamos National
Laboratory, Los Alamos, NM 87545.

\pacs{PACS Numbers: 71.27.+a, 75.20.Hr, 75.30.Mb, 75.40.Mg}

\newpage
\thispagestyle{empty}

\section{Introduction}

It is well known that a magnetic Anderson impurity in a metal is 
screened by conduction electrons, and
a singlet is formed at low temperatures. However, very little
attention has been given to the nature of the
ground state when a magnetic impurity is in an insulator
or a semiconductor. That is the subject of this paper.
One might think that a magnetic impurity in an
insulator or a semiconductor will not be screened because there is
a gap in the density of states. However, the problem
is somewhat more complicated than this simple expectation.

Let us briefly review the previous work in this field.
Withoff and
Fradkin \cite{Fradkin} used mean field theory to consider a
Kondo impurity in a system where the density of states 
goes to zero at the Fermi energy with power law behavior,
i.e., $\rho(\epsilon)\sim|\epsilon|^{r}$ where $|\epsilon|<D$ ($D$
is the bandwidth) and $r$ = 0, 1/2, 1, or 2. In this gapless
situation they found that
the Kondo impurity became a singlet only if the spin exchange
$J>J_{c}$, where the critical coupling $J_{c}$ was a 
function of the power $r$.
Later authors considered a finite energy gap $\Delta$ in the 
density of states.
Takegahara {\it et al.} \cite{symjap} used Wilson's numerical
renormalization group to argue that the symmetric Anderson
impurity is always a magnetic multiplet
for any finite gap $\Delta$, while the asymmetric Anderson impurity
\cite{asymjap} has a critical value of the gap $\Delta_{c}$ 
such that
the impurity has a magnetic ground state if the gap is too 
big, i.e., if $\Delta > \Delta_{c}$. 
Ogura and Saso \cite{saso} found no such qualitative difference 
between the symmetric and asymmetric impurities. They have
examined the problem using poor man's scaling, the 1/N 
expansion, the
non-crossing approximation, and quantum Monte Carlo. They
argued that both the symmetric and asymmetric Anderson
impurity remain magnetic if the semiconducting
gap $\Delta$ is ``large enough''; otherwise the impurity is screened
and forms a singlet at low temperatures. ``Large enough'' means that
$\Delta \ge \alpha T_{K}$, where $T_{K}$ is the Kondo temperature
and the value of $\alpha$ varies between 0.4 and 2.0, depending
on the parameters and the calculational approach. Since there is
a gap, it is somewhat artificial to define a Kondo temperature, but
Ogura and Saso define it by $T_{K}=D\exp(-1/J\rho_{o})$, where
$\rho_{o}$ is the flat density of states of the semiconductor outside
the gap. 

All of these considerations focus on whether or not the impurity
is screened and becomes a singlet at low temperatures. This tacitly 
assumes
that the magnetic moment is localized at the impurity site. This is
not necessarily true. As we shall see, if there is a large
hybridization between the localized orbital and the conduction
orbital on the impurity site, the spin and charge are not
localized at the impurity site, but are extended throughout the 
system. The importance of hybridization 
is well known in semiconductors. After all, it is hybridization 
that allows ordinary donor and acceptor impurities to 
contribute carriers to a semiconductor. Within the Hartree--Fock
approximation, Haldane and Anderson \cite{Anderson2} found that
hybridization between the d-orbitals of a transition metal
impurity and the valence band electrons of the semiconductor
allowed nominally different charge states to exist as states in
the gap.

In this paper we place an Anderson impurity in the middle of
a one--dimensional semiconductor.
We use the $U=0$ one--dimensional Anderson Hamiltonian at half 
filling as the semiconducting host, 
since it has a hybridization gap. By varying the
hybridization $V$, we can control 
the size of the semiconducting gap. We consider chains
with an odd number of sites, and we place the substitutional
Anderson impurity in the middle of the chain. 
The impurity has a positive Coulomb repulsion $U_{0}>0$ and
a positive hybridization $V_{0}$. In section II we present the
Hamiltonian, which we study using the density matrix renormalization
group approach.\cite{White} The advantage of 
this technique is that it is done
in real space so that we can evaluate how correlation functions 
vary with distance as well as examine the spin and 
charge densities as a function of position.

In section IIIA we discuss
the half--filled case; we find that the ground state always has 
total spin $S=0$. We also study the spin-spin correlation functions
to see if the impurity spin persists spatially.
We find that the spin correlations decay exponentially
with distance due to the presence of the gap. 
We compare these results to the
metallic case in which we place the Anderson impurity in a free 
electron
host with the same number of sites as in the semiconducting case.
In the metallic case we find that the spin correlations decay 
more slowly with distance. We would also like to
plot the spin as a function of position. However, the
half--filled case has a singlet ground state with
$S_{i}=0$ on each site. So in section IIIB, we consider
the doped case in which we add a hole to the system; this makes
the total spin $S=1/2$. In the doped semiconducting case
we find two different regimes. When the hybridization of the impurity
site is small, the spin and charge of the
hole are localized near the impurity. When
the hybridization is large, the spin and charge
of the hole are delocalized and reside in the host away from the 
impurity; a singlet is formed on the impurity site.
Thus, in the semiconducting case, the spin and the hole go together 
and have similar spatial distributions. In contrast,
in the doped metallic case, the charge density of the
hole is delocalized for all values of the impurity hybridization
$V_{0}$. However, the spin density of the hole is localized
near the impurity for small values of the hybridization due to
finite size effects. For large
values of $V_{0}$, finite size effects no longer dominate; 
a singlet forms on the impurity site, and the
hole and spin densities are delocalized. 

When more than one hole is added to the semiconductor, the 
additional holes avoid the impurity and spread throughout the
lattice. 
%The impurity acts like a very high barrier in a double well
%potential, and divides the lattice in two.
We show that there is
an analogy between the barrier in a double well potential and
the impurity in the semiconductor doped away from half--filling.
Finally in section
IIIc, we discuss how the chemical potential varies with filling.
For small $V_{0}$, we find states lying in the middle of the gap.
As $V_{0}$ increases, these states move toward the edges of the
gap. 

\section{\label{PERANDSEC} The 1D Anderson Hamiltonian 
with an Impurity}

The standard one--dimensional periodic Anderson lattice has spin-1/2
conduction electrons that hop from site to site. Each site has
a localized f-orbital with a Coulomb repulsion $U(i)$ and a 
hybridization
$V(i)$ between the conduction orbital and the f-orbital.
The Hamiltonian is given by:

\begin{equation}
H=-t\sum_{i \sigma} (c^{\dagger}_{i \sigma}   c_{i+1 \sigma} +
                     c^{\dagger}_{i+1 \sigma} c_{i\sigma} )
  + \sum_{i \sigma}\varepsilon_{f}(i) n^{f}_{i \sigma} 
  + \sum_{i} U(i) n^{f}_{i \uparrow}n^{f}_{i \downarrow}
  + \sum_{i \sigma}V(i) (c^{\dagger}_{i \sigma} f_{i \sigma} + 
                      f^{\dagger}_{i \sigma}  c_{i \sigma} )     
\label{eq:Hamiltonian}
\end{equation}
where $c^{\dagger}_{i \sigma}$ and $c_{i\sigma}$ create and 
annihilate conduction 
electrons with spin $\sigma$ at lattice site $i$, 
and  $f^{\dagger}_{i \sigma}$ and $f_{i \sigma}$ 
create and annihilate local $f$--electrons. 
Here $t$ is the hopping matrix element for conduction 
electrons between neighboring sites, 
$\varepsilon_{f}(i)$ is the energy of the 
localized $f$--orbital at site $i$,  
$U(i)$ is the on--site Coulomb
repulsion of the $f$--electrons and $V(i)$ is the 
on--site hybridization matrix element 
between electrons in the $f$--orbitals and the conduction band.
For simplicity, we neglect orbital degeneracy. 
We denote the number
of electrons by $N_{el}$, and $L$ is the number of sites 
in the lattice. 
$t$, $U(i)$, $V(i)$, and $\varepsilon_{f}(i)$ are taken to be 
real numbers.

In order to find the semiconducting gap $\Delta$, we note that
the uniform periodic Anderson Hamiltonian with $U(i)=0$ and
$V(i)=V$ independent of $i$ can be diagonalized exactly in $k$ 
space.  We obtain
two hybridized bands with energies $\lambda_{k}^{\pm}$:
\begin{equation}
\lambda_{k}^{\pm}=\frac{1}{2}\left[(\varepsilon_{f}-2t\cos(ka)) \pm
\sqrt{(\varepsilon_{f}+2t\cos(ka))^{2}+4V^{2}}\right]
\end{equation}
where $a$ is the lattice constant.
When there are two electrons per unit cell, the lower band 
is full while the upper one
is empty. Thus the system is insulating when $N_{el}=2L$.
The size of the band gap is given by 
\begin{equation}
\Delta=\lambda_{k=0}^{+}-\lambda_{k=\pi/a}^{-}=2\sqrt{t^{2} 
+ V^{2}} - 2t
\label{eq:gap}
\end{equation}

In this paper we consider chains with an odd 
number of sites and we place the impurity in the center of the chain.
We denote the site at the middle by $i=0$. (For example, for a 25
site chain, $i$ runs from --12 to +12.) For $i \ne 0$, we set 
$U(i)=0$ and $V(i)=V$. At half filling ($N_{el}=2L$), 
these sites represent the
semiconducting host since there is a hybridization gap between
the conduction band and the flat f-band. We set $t=1$;
this sets the energy scale. By varying the
value of $V$ we can control the size of the gap.
The impurity site at $i=0$
has $U(0) = U_{0}$, $V(0) = V_{0}$, and 
$\varepsilon_{f}(0)=\varepsilon_{f0}$.

We primarily consider the symmetric case, for which 
$\varepsilon_{f}(i)=-U(i)/2$.
With this choice, the half--filled case has particle--hole symmetry
and there is an SU(2) charge pseudospin symmetry.\cite{Nishino}
The components of the pseudospin operator $\vec{I}$ are given by:
\begin{eqnarray}
\nonumber  I_{z} & = & \frac{1}{2} \sum_{i} (c^{\dagger}_{i \uparrow}
c_{i \uparrow}+
            c^{\dagger}_{i \downarrow}c_{i \downarrow}+
            f^{\dagger}_{i \uparrow}f_{i \uparrow}+
            f^{\dagger}_{i \downarrow}f_{i \downarrow} -2) \\
            I_{+} & = & \sum_{i} (-1)^{i} (c^{\dagger}_{i \uparrow}
                                       c^{\dagger}_{i \downarrow}-
             f^{\dagger}_{i \uparrow}f^{\dagger}_{i \downarrow}) \\
\nonumber  I_{-} & = & \sum_{i} (-1)^{i} 
(c_{i \downarrow}c_{i \uparrow}-
        f_{i \downarrow}f_{i \uparrow})  
\end{eqnarray}
The z--component of the pseudospin is equal to $(N_{el}/2) - L$. 
Note that
half--filling corresponds to $I_{z}=0$. An $I_{z}=1$ state can be 
achieved by adding two electrons. All the energy 
eigenstates of the symmetric Anderson model have a 
definite value of S and I. At half filling
the ground state is a singlet both in spin and pseudospin space 
$(S=0, I=0)$, even when an impurity is present.

Since Takegahara {\it et al.} \cite{symjap,asymjap} have emphasized
the difference between a symmetric and an asymmetric 
Anderson impurity,
we also consider an asymmetric Anderson impurity in a symmetric
$U=0$ Anderson lattice. We find no qualitative difference between
the symmetric case and the asymmetric case.

We compare our results for the semiconductor to the
metallic case in which we place the Anderson impurity in the center
of a one--dimensional free electron
host with the same number of sites as in the semiconducting case.
The Hamiltonian for the metallic case is given by
\begin{equation}
H=H_{o}+H_{imp}
\label{eq:metimpham}
\end{equation}
where the free electron tight--binding Hamiltonian $H_{o}$ is
given by
\begin{equation}
H_{o}=-t\sum_{i \sigma} (c^{\dagger}_{i \sigma}   c_{i+1 \sigma} +
                     c^{\dagger}_{i+1 \sigma} c_{i\sigma} )
\label{eq:metham}
\end{equation}
and the impurity Hamiltonian $H_{imp}$ is given by
\begin{equation}
H_{imp}=\sum_{\sigma}\varepsilon_{f}(0) n^{f}_{0 \sigma}
  + U(0) n^{f}_{0 \uparrow}n^{f}_{0 \downarrow}
  + \sum_{\sigma}V(0) (c^{\dagger}_{0 \sigma} f_{0 \sigma} +
                      f^{\dagger}_{0 \sigma}  c_{0 \sigma} )
\label{eq:impham}
\end{equation}

For future reference, it is convenient to define the effective
spin exchange coupling $J_{eff}$. This comes from considering
a single Anderson impurity in a metal.
When the hybridization term is small
($\pi \rho(\varepsilon_{f}+U)V^{2}/(\epsilon_{f}+U) \ll 1$ and
$\pi \rho(\varepsilon_{f}) V^{2}/\epsilon_{f} \ll 1 \,$), 
the single impurity Anderson Hamiltonian can be mapped into
the Kondo Hamiltonian \cite{Schrieffer}
\begin{equation}
H_{K}= J_{eff}\vec{S}^{f} \cdot \vec{s}^{c}_{0}
\end{equation}
where $\vec{s}^{c}_{0}$ is the spin density of the 
conduction electrons at the impurity site and
$J_{eff}$ is given by the Schrieffer-Wolff transformation: 
\cite{Schrieffer}
\begin{equation}
J_{eff}=-\frac{2|V|^{2}U}{\epsilon_{f}(\epsilon_{f}+U)} 
\label{eq:Jeff}
\end{equation}
Note that for the symmetric case where $\epsilon_{f}=-U/2$, 
$J_{eff}=8V^{2}/U$.

We use the density matrix renormalization
group (DMRG) method \cite{White} to calculate the ground state
as well as to determine the spin--spin correlation functions,
and the spin and charge densities in the ground state.
The DMRG appoach is a real space technique which has proven
to be remarkably accurate for one dimensional systems
such as the Kondo and Anderson lattices.\cite{guerrero,metal}
We use the finite system method with open boundary 
conditions in which
there is no hopping past the ends of the chain. We study lattices
with 25 sites, keeping up to 120 states with typical truncation
errors of order $10^{-7}$ for the semiconductor. For the metal
we keep up to 130 states with typical truncation errors of order
$10^{-6}$.

\section{Results}
\subsection{Half--Filled Case}
In this section we consider the half--filled case. In particular
we want to see how far the impurity spin persists by studying
the spin-spin correlation functions. We compare the gapped
semiconducting case with the gapless metallic case. We find
that the spin correlations fall off much faster in the presence
of a gap.

To model the semiconductor, we use Anderson lattice chains with
25 sites and $N_{el}=2L=50$ (2 electrons per site or one 
electron per orbital). The Hamiltonian is given by eq. 
(\ref{eq:Hamiltonian}) with $U=\epsilon_{f}=0$ for $i \ne 0$,
and $U_{0}>0$ and $ \epsilon_{f0}=-U_{0}/2$ for the 
symmetric impurity 
site with $i=0$. We set $t=1$. For $V=1$, the semiconducting gap
$\Delta \sim 0.83$ for a 25 site chain with open boundary 
conditions.\cite{gap} 
We initially set $U_{0}=8$ and $V_{0}=1$ at the impurity site.

We model the half--filled metal with the tight--binding 
Hamiltonian of eqs.
(\ref{eq:metimpham},\ref{eq:metham},\ref{eq:impham}). 
In this case $N_{el}=26$.
We set $t=1$ and place the impurity at the center at site $i=0$
with the same values of $U_{0}$ and $V_{0}$ that we used for the
semiconductor.

For both the metallic and semiconducting cases, we find 
that the ground
state has $I=0$ and $S=0$. We show the spin-spin 
correlation functions for both cases in Fig. \ref{fig1}. We plot the
correlations $<S_{z}^{c}(R)S_{z}^{f}(0)>$
between the z--component of the f--spin at the impurity site and the 
z--component of the spin of 
the conduction electrons in the lattice versus the distance $R$
from the impurity.
Fig. \ref{fig1}a is a linear plot. The metallic case clearly shows
Friedel oscillations. The bias of the data for the metal
toward negative values of
the correlations is a finite size effect; we show in the appendix
that the leading term in perturbation theory for 
$<S_{z}^{c}(R)S_{z}^{f}(0)>$ goes as $-1/L$.
Fig. \ref{fig1}b is a linear-log plot of the absolute value 
of the same correlations; we have removed the oscillations
by plotting every other point. 
Here we see that in the metallic case
the correlations between the impurity f-spin and the conduction
spins, which are responsible for the compensation of the magnetic
moment, decay very slowly. In contrast, in the
semiconducting case the decay is much faster due to the presence
of a gap in the excitation spectrum. If assume that the correlations
fall off exponentially and fit the plots in Fig. \ref{fig1}b
to the form $\exp(-R/\xi)$,
then the correlation length $\xi/a\sim 1.9$ for the metal and 
$\xi/a\sim 0.49$ for the semiconductor.
%({\it sorensen and affleck: they determine $\xi_{K}$ by fitting
%$S_{z}$ to a scaling function that goes as $L/\xi_{K}$.
%The fit to determine $\xi$ is fine for Figs. 1b and 2.
%To see this, just draw a straight line through the data in Figs. 1b 
%and 2.
%I used the xmgr regression package that was on the U. of Florida
%machines. It doesn't work so good here. 
%So I took the natural logarithm
%of the y-values (correlation fcns),
%and used xvgr linear regression to fit the data to a straight line.
%1/slope=$\xi$.})

We attribute the fact that the spin correlations 
decay much faster in the semiconductor than in the metal due to the
presence of semiconducting gap. To check this, we can change the
size of the gap in the semiconductor by varying $V$. This should
change the spin-spin correlation length. This is confirmed in
Fig. \ref{fig2} where we plot the spin-spin correlation functions 
for the semiconducting case for 2 different values of
$V$ in a linear-log plot with $U_{0}=8$. 
For $V=0.4$, $\Delta\sim 0.16$, and $\xi/a\sim 0.95$;
for $V=1$, $\Delta\sim 0.83$, and $\xi/a\sim 0.49$.\cite{gap}
Thus we see that as $V$ and hence the gap $\Delta$ increase,
the correlation length decreases as expected.
%i.e., the correlation length varies inversely with the gap.

We varied the hybridization $V_{0}$ at the impurity site 
to study the effect on the spin--spin correlation 
functions at half--filling for both the metallic and the 
semiconducting 
cases. In the semiconducting case, we find that changing $V_{0}$
from 0.5 to 10, while keeping the semiconducting gap 
constant, does not change the qualitative 
behavior of the correlation functions. Similarly, in the metallic
case we find that changing $V_{0}$ from 0.1 to 10 does not change
the qualitative behavior of the correlation functions.

We also examined an asymmetric Anderson impurity in a 
semiconductor at half--filling with $L=25$, $N_{el}=50$,
$t=1$, $V=1$, $U=0$, $U_{0}=16$, and $\varepsilon_{f0}=-0.5$.
We found no qualitative difference in the spin--spin
correlation functions as $V_{0}$ varied between 0.1 and
10. In fact, the the behavior of the correlation functions
was very similar to that found in the symmetric case. 

\subsection{Doped Case}
%\subsubsection{One Hole}

If we plot the spin and charge density as a function of
position in the half--filled case, we find that 
the spin is zero and the charge density is 2 at every site.
In order to obtain more interesting positional information, we
dope our system of 25 sites by adding a hole.
We put $N_{el}=49$ in the semiconducting case and $N_{el}=25$ in the 
metallic case. The total spin in the ground state is $1/2$ 
since there is
an odd number of electrons. This corresponds to a quasiparticle
excitation of
the half--filled system. Again, we set $t=1, V=1$, and $U=0$ 
in the host.
By fixing $V$, we set the value of the gap 
$\Delta=0.83$ in the semiconducting case. We initially consider
a symmetric Anderson impurity with $U_{0}=8$ and we vary $V_{0}$.
One can think of changing
$V_{0}$ as corresponding to changing the effective Kondo coupling
$J_{eff}=8V_{0}^{2}/U_{0}$, although this picture is only valid 
for small $V_{0}$.

We study the hole density and the spin density versus site.  
(Here the hole density refers to the number of holes per site 
measured relative to the half--filled case.)
Naively, we can think of  two possibilities:
the hole can be localized in the impurity site or it can be 
delocalized and spread out in the rest of the chain. 
The results for the semiconducting case appear in Fig. \ref{fig3}.
We can clearly identify two regimes in the semiconductor:

i) Large $V_{0}$ : the hole and the spin density are delocalized 
and reside 
   in the host. On the impurity site the hole and spin density are
   zero, which means that the impurity has
   an f-electron and a conduction electron combined in a singlet 
state. 

ii) Small $V_{0}$ : the spin density is localized at 
the impurity site
   while the hole density is localized on the impurity site and its
   nearest neighbors.

To understand what determines whether or not the hole is
localized, we must compare the energy of adding a delocalized
hole to the host versus the energy of adding a hole to the impurity.
Removing an electron from the semiconductor costs roughly half the
gap ($\Delta/2$). To estimate the energy of putting a hole on 
the impurity,
we consider the Hamiltonian of an isolated single site 
Anderson impurity:

\begin{equation}
H_{0}=U_{0}n^{f}_{\uparrow}n^{f}_{\downarrow}-
\frac{U_{0}}{2}(n^{f}_{\uparrow}+
n^{f}_{\downarrow}) + V_{0} \sum_{\sigma}
(c^{\dagger}_{\sigma}f_{\sigma} +
f^{\dagger}_{\sigma}c_{\sigma})
\end{equation}
Since this is a symmetric impurity, we can classify the 
states by their value of $S$ and $I$. We find
\begin{itemize}
\item 2 \, ($S=0 ,\,   I=0 $) \, \, \, \, \, states with \, \, 
$E^{00}_{\pm}=-\frac{U_{0}}{4}\, \pm \,
\frac{1}{2}\sqrt{\frac{U_{0}^{2}}{4}+16V_{0}^{2}}$
\item 2 \, ($S=1/2 ,\, I=1/2$) \, states with \, \, 
$E^{\frac{1}{2}\frac{1}{2}}_{\pm}=-\frac{U_{0}}{4}\, 
\pm \,\frac{1}{2}
\sqrt{\frac{U_{0}^{2}}{4}+4V_{0}^{2}}$
\item 1 \, ($S=1 ,\, I=0$) \, \, \, \, \,  state 
with \, \, $E^{10}= -\frac{U_{0}}{2}$     
\item 1 \, ($S=0 ,\, I=1$) \, \, \, \, \,  state 
with \, \, $E^{01}=0$
\end{itemize}
Here the first superscript of $E$ indicates the value of $S$ 
and the second one 
indicates the value of $I$ in that state. For a single impurity, the 
$I=0$ states have two electrons,
while the $I=1/2$ states can have 3 electrons ($I_{z}=+1/2$) or 
one electron ($I_{z}=-1/2$).
 
The lowest energy state is the lowest state with $S=0,\,I=0$ for 
any choice
of parameters. In general we find
\begin{equation}
E^{00}_{-} \, \, < \, \, E^{\frac{1}{2}
\frac{1}{2}}_{-} \, \, < \, \, 
E^{10}    \, \, < \, \, E^{01} < \, \, 
E^{\frac{1}{2}\frac{1}{2}}_{+} \, \, < \, \, E^{00}_{+} 
\end{equation}
The difference in energy $\Delta E$ between the two lowest states is
\begin{equation}
\Delta E \equiv E^{\frac{1}{2}\frac{1}{2}}_{-}\, - \, E^{00}_{-}
%=\frac{1}{2} \sqrt{\frac{U_{0}^{2}}{4}+16V_{0}^{2}}-
%\frac{1}{2} \sqrt{\frac{U_{0}^{2}}{4}+4V_{0}^{2}}
\end{equation}
For large $V_{0}$ ($16V_{0}^{2} \gg U_{0}^{2}/4$) one gets 
$\Delta E \sim V_{0}$
and for small $V_{0}$, $\Delta E \sim 6V_{0}^{2}/U$. This difference
$\Delta E$ represents the energy 
cost to put the hole at  the impurity site. If, on the 
other hand, the 
hole goes to the host, the energy cost is roughly equal to half the 
gap ($\Delta/2$). 
Therefore, when $\Delta E \, < \, \Delta/2$, 
the hole should go to the impurity
site, meaning that the impurity should be in the $S=1/2, \, I=1/2$ 
state with $S_{z}=1/2$ and $I_{z}=-1/2$. 
According to this criteria, the crossover should occur when
$\Delta E=\Delta/2$. For the values of the parameters that we use,
this crossover corresponds to 
$V_{0} \sim 1.25$. For $V_{0}$ less than $1.25$, 
the hole and the spin density should be localized 
at the impurity site because the 
$S=1/2, \, I=1/2$ state is more favorable, but when $V_{0}$ is 
greater than $1.25$, the impurity should 
be in the $S=0, \, I=0$ state and the hole 
and  the spin density should be spread out over the lattice. 
This is consistent with 
the numerical results, since for $V_{0}=1$ the hole is localized 
while for $V_{0}=2$ it is spread out over the lattice.
In the crossover region ($1<V_{0}<2$) the values of the
hole and spin densities on the
impurity site are intermediate between those found for
$V_{0}=1$ and $V_{0}=2$. 
%For example, for $V_{0}=1.5$ and $L=25$,
%he hole density on the impurity site is 0.09, and the total
%pin density on the impurity site is $S_{z}(0)=0.1$. 
However, this gradual crossover may be a finite size effect
since we have only looked at lattices up to 25 sites long.

Fig. \ref{fig3} shows that for small $V_{0}$ ($V_{0}<1.25$),
the hole density is localized on the impurity as well
as its nearest neighbor sites. This can be understood as follows: 
the hole density likes to be
localized at the impurity according to the criteria explained above.
However, electrons on neighboring sites optimize their
kinetic energy by hopping into the hole on the impurity site.
Thus the hole spreads to the
two nearest neighbor sites of the impurity. This is confirmed
in Fig. \ref{ke} which shows of the kinetic energy
of the bonds between sites as a function of position.
Notice that the bonds connecting
the impurity site have the largest magnitude of
the kinetic energy for the smallest values of $V_{0}$.
We have looked at the conduction and f--electron density on
the sites neighboring the impurity. We find that when
the hole resides on these sites,
it is primarily in the f-orbital where the energy cost is zero
because $\varepsilon_{f}(i\neq 0)=-U/2=0$
on these sites. The electrons on these sites are
in the conduction orbitals where they can take advantage of the
kinetic energy.

Let us discuss what dictates where the spin of the hole resides.
When the impurity hybridization $V_{0}$ is large,
a singlet forms between the conduction spin and the f-spin
on the impurity site.
Thus the impurity has no net spin, and the spin of the hole 
resides in the host. Will it reside primarily in the f-orbitals
or in the conduction orbitals? To answer this, we note that
if the hybridization $V$ of the host is not
too large, then optimizing
the kinetic energy of the conduction electrons dominates over
optimizing the hybridization energy of the host. In order to
allow both up and down spin conduction electrons to hop
freely from site to site, the average spin of the conduction
electrons on each site is zero. Thus the spin of the hole 
must be spread primarily over the f-orbitals of the host lattice.

On the other hand, when the impurity
hybridization $V_{0}$ is small, the 
hybridization on the host sites has priority. 
This favors the formation of singlets on the host sites.
As a result, the spin of the hole will be localized
primarily on the impurity site. In order to minimize
the kinetic energy of the conduction electrons, the
spin will primarily reside in the f--orbital of the impurity. 
This occurs at the expense of the hybridization energy of
the impurity, but that is permissible since this is the smallest
energy in the problem.
The arguments of the last two paragraphs indicate that 
the spin of the hole will
be primarily in the f-orbitals for the range of parameters
that we studied. This is shown in Fig. \ref{spinsum}.

We now consider the metallic case.
It is easy to compare the energy of adding the hole to the host
versus the finite energy $\Delta E$ of localizing the hole on the 
impurity. For a metallic host, the chemical potential is zero
at half--filling, and there
is no energy cost in adding a delocalized hole to the metal.
Thus, one expects that the hole will always
be spread out and extended throughout the metal.
In Fig. \ref{metalhole} we show the numerical results.
We see that the hole density behaves as expected: it is 
spread out over the 
lattice for every set of parameters that we examined. 
For large $V_{0}$, the large
on--site hybridization favors a singlet state at the
impurity and the spin density is spread
out over the lattice. In this case the spin is in the conduction
spins because there are no f--orbitals in the metallic host.
However, for small values of $V_{0}$,
the spin density becomes 
localized at the impurity. We attribute this to finite size effects
since we expect a singlet at the impurity site in an 
infinite metallic lattice. We can understand how finite size effects 
affect the behavior of the spin density in the following way.
If the size of the lattice is such that the spacing 
between discrete energy levels of the metallic host
becomes comparable to or larger than $J_{eff}$, then the 
exchange interaction
is too weak to mix the noninteracting conduction energy levels enough
to form a singlet with the f-spin. In this case, 
there will be a magnetic
moment on the impurity site. We can check this explanation 
by comparing $J_{eff}$ with the energy level spacing. 
For a 25 site metallic lattice with open boundary conditions, 
the typical energy level spacing is 0.24. We can compare this
with $V_{0}=0.1$ which has $J_{eff}=0.01$, and with
$V_{0}=1$ which has $J_{eff}=1$. As we can see in
Fig. \ref{metalhole}b, these two cases have a local magnetic moment.
The $V_{0}=2$ case is borderline and has a small magnetic moment
at the impurity site. The influence of finite size effects 
can be seen
by putting a symmetric Anderson impurity with $V_{0}=2$ and 
$U=8$ in the middle of a 7 site lattice.
We find that the spin density on the impurity site is roughly 
twice that found for a 25 site lattice (see Fig. \ref{metalhole}b).
Finally we note that finite size effects do not affect our
results for a semiconductor because the semiconducting gap
is much larger than the energy level spacing. 
For example, a 25 site symmetric Anderson lattice with open
boundary conditions with $U=0$ and $V=1$ has
a typical energy level spacing of 0.01 which is much smaller
than the semiconducting gap of $\Delta=0.83$. Similarly
if $V$ is changed to 0.4, the energy level spacing is still
approximately 0.01 which is much smaller than $\Delta=0.15$.

We also examined the asymmetric Anderson impurity in a semiconductor
doped with either one hole or one particle with $L=25$, $N_{el}=49$, 
$t=1$, $V=1$, $U_{0}=16$,
$\varepsilon_{f0}=-0.5$. The behavior of the spin and
charge densities at small $V_{0}$ ($V_{0}=0.1$)
and at large $V_{0}$ ($V_{0}=10$) is very similar to that found
for the symmetric Anderson impurity. 

%\subsubsection{Two Holes}
We will devote the rest of this section to discussing the fact
that the impurity provides a large potential barrier and effectively
divides the lattice in two as the system is doped away from 
half--filling.
As a result, we can think of the semiconductor as a symmetric
double well potential. There are several examples of where this 
occurs. For example, consider what happens when we add two holes
to a half--filled semiconductor with a symmetric Anderson
impurity. As before, we set $t=1$, $V=1$, and $U=0$ in the host.
We place the Anderson impurity in the center of a 25 site lattice
with $U_{0}=8$, and we vary $V_{0}$. Adding two holes corresponds
to $N_{el}=48$. 
For small $V_{0}$, the ground state consists of two states
which are degenerate within the accuracy of our 
calculation.\cite{error}
In one state the system is a singlet and in the other it is a 
triplet.
This near degeneracy is not the result of finite size effects or
boundary conditions since we find this
degeneracy for smaller lattice sizes as well as for the case
of periodic boundary conditions.
By examining the hole density versus site as shown in 
Fig. \ref{fig:2holes},
we find that one hole is localized on the impurity site
and its two nearest neighbors, while the other hole is spread
over the lattice. We cannot put two holes on the impurity
because that would involve removing the f--electron from the
impurity which would cost an energy of $U_{0}/2$. As a result,
the additional hole avoids the impurity and its two nearest
neighbors, and spreads over the host. It resides primarily in
the f--orbitals where the energy $\varepsilon_{f}(i)=-U/2=0$.

The impurity site with
the hole localized in its vicinity 
acts like an nearly infinite potential barrier to the second
hole and effectively divides the lattice in two.
Thus, the energy associated with adding the second hole 
should be equal to that of adding a hole to
a 22--site semiconductor ($t=1$, $V=1$, and $U=0$) with no
impurity but with
a break ($t=0$) in the middle. We can think of this semiconductor
as a symmetric double well potential with a nearly infinite 
barrier. Each
potential well corresponds to an 11--site semiconductor.
Putting a hole on the right side or on the left side or taking
a linear combination of these two cases results in states
whose energies are nearly degenerate. (This explains the degeneracy
of the ground state.\cite{degen}) Within the limits of our accuracy,
we find that the energy
to put one hole in an 11--site half--filled semiconductor with
no impurity is indeed equal to the energy to add a second hole
to a 25--site semiconductor with an impurity in the middle.
Our double well scenario is further confirmed by the fact that
the energy associated with adding the second hole is the
same for $V_{0}=0.1$ and $V_{0}=1$ within our accuracy.
This is consistent with having a very high barrier for both
cases.

On the other hand,
for large $V_{0}$, we find that both holes go
into the host lattice and a singlet forms between the
f--electron and the conduction electron on the impurity site.
This singlet acts like a potential barrier, but since having
two holes on one side of the barrier versus having one hole on
each side are not degenerate states, the ground state is 
nondegenerate. In fact,
the ground state of the whole system is a singlet.
However, if we keep $V_{0}$ large but have one hole rather than 
two holes, the singlet on the impurity
acts like a very high barrier which divides the wavefunction for
the hole into two pieces (see Fig. \ref{fig3}). Since 
having the hole on
one side of the impurity versus the other are nearly
degenerate configurations, the ground state is nearly degenerate
and both states have spin--1/2.

It is easy to generalize these trends to cases where more than one
hole is doped into a half--filled system. For small $V_{0}$, the
first hole resides in the vicinity of the impurity, 
and the additional holes avoid
the impurity and are extended throughout the lattice. For
large $V_{0}$, a singlet forms on the impurity site; 
the holes avoid the impurity and are extended throughout
the lattice. For both large and small $V_{0}$, the ground state 
is nearly degenerate when the
number of extended holes is odd. For example, when there are four
holes and $V_{0}$ is small, one hole resides in the vicinity
of the impurity; the remaining 3 holes are spread over the
rest of the lattice, and the ground state
is nearly degenerate.

\subsection{Chemical Potential Versus Filling}
In this section we study how the chemical potential varies with
electron filling. As in the previous sections, we consider a
symmetric Anderson lattice with
$t=1$, $U=8$, $V=1$, and $L=25$ with the impurity site
in the middle of the lattice. We define the chemical potential by
\begin{equation}
\mu(N)=E(N)-E(N-1)
\end{equation}
where $E(N)$ is the ground state energy with $N$ electrons. Our 
results
are shown in Fig. \ref{fig:chempot}. When the impurity is absent,
there is a jump in the chemical potential that is centered about
half--filling ($N=50$). This is the quasiparticle gap. From Fig.
\ref{fig:chempot}, we see that there are states in the gap
for small $V_{0}$. The chemical potential
of these midgap states corresponds to the energy of adding
a particle or a hole to the half--filled system.
These midgap states move
to the edges of the gap as $V_{0}$ increases. Indeed, they appear
to merge with the gap edges for $V_{0}\geq 2$. The fact that
the impurity does not seem to affect the chemical potential
for large values of $V_{0}$ is consistent with the delocalization
of the hole density and its spin which we saw in the last section.
The presence of midgap states for small values of $V_{0}$ is
consistent with the localization of the hole and its spin. To see
this, suppose that $V_{0}\ll V$. Then the f--orbital on the
impurity decouples from the rest of the lattice.
In addition the large hybridization $V$
favors having one conduction electron and one f--electron
on each of the host sites. As a result, when we put 
0, 1, or 2 conduction
electrons on the impurity site, the associated particles or holes 
will be localized in the vicinity of the impurity, and 
the energies of these states will be nearly degenerate. This 
means that
the chemical potential corresponding to adding a particle or a
hole to a half--filled system is close to zero. This is indeed what 
we see for $V_{0}=0.1$.

\section{\label{CONCLUSION} Conclusions}
In this paper we have studied an Anderson impurity in a 
one--dimensional semiconductor. Although we primarily
concentrated on a symmetric Anderson impurity, we found no 
qualitative 
difference in behavior between
an asymmetric impurity in the mixed valence regime and a symmetric
impurity in the Kondo regime.
In the undoped half--filled case we found spin-spin correlation 
functions
that decay rapidly with distance due to the gap in the excitation
spectrum. This is in contrast with the metallic case in which a much
slower decay is seen.

Because DMRG is a real
space technique, we were able to go beyond the question
of whether or not the magnetic impurity is screened in 
the presence of a gap in the density of states. In the
case of doping with an $S=1/2$ hole, we found that
a large on--site hybridization $V_{0}$ led
to the formation of a singlet on the impurity site and the 
delocalization
of the spin and charge density throughout the lattice.
For small $V_{0}$, the magnetic moment of the hole
was localized on the impurity site, and the charge
density was concentrated on the impurity and its nearest
neighbors. The criteria for defining these two regimes
was whether it costs more energy to put the hole on
the impurity site or to spread it throughout the lattice.
This is different from the criteria used by Ogura and
Saso \cite{saso} who found that the impurity remained
a magnetic multiplet if the semiconducting gap $\Delta$
was greater than some fraction of the Kondo temperature
$T_{K}$. It is somewhat artificial to define a Kondo
temperature since there is a gap at the Fermi energy,
but let us define it by $T_{K}=D\exp(-1/J_{eff}\rho_{o})$, where
$\rho_{o}=2/\pi^{2}t$ is the density of states at the Fermi
energy for free electrons with open boundary conditions, 
and $D=4t$
is an estimate of the conduction electron bandwidth. Then we can
compare our results with those of Ogura and Saso.\cite{saso}
We find that the charge and spin density of the hole are 
localized for $\Delta > T_{K}$, and are extended for
$\Delta < T_{K}$. This agrees qualitatively with Ogura
and Saso.\cite{saso}

We compared our semiconducting results with those of a metal.
When we put a hole into the half--filled metal, 
we find that a singlet forms if $V_{0}$ is large.
For $V_{0}$ small, the magnetic moment of the hole is
localized on the f-orbital of the impurity due to 
finite size effects.
The charge density of the hole is extended for all
values of $V_{0}$ since it always costs less energy
to put the hole in an extended wavefunction than
to localize it in the vicinity the impurity.

We found that the impurity in a semiconductor doped away from
half--filling acts like a barrier in a symmetric double well 
potential.
When $V_{0}$ is large, a singlet forms on the impurity site. This
singlet acts like a barrier that divides the lattice in two. 
The holes in the system avoid the impurity and spread over the 
rest of
the lattice. When $V_{0}$ is small, the first hole goes onto the
impurity which acts like a barrier and divides the lattice for the
rest of the holes. These additional holes spread over the two halves
of the lattice. When the number of delocalized holes is odd, the
ground state is nearly degenerate for both large and small values
of $V_{0}$.

Finally, we studied the chemical potential as a function of 
electron filling.
We found that midgap states appear for small values of $V_{0}$
and correspond to localization of a hole or particle on the impurity 
site. As $V_{0}$ increases, these midgap states
move towards the edges of the gap which is associated with the
delocalization of the hole.

It may be possible to look for some of the effects we have described
in dilute magnetic semiconductors \cite{furdyna}. For example,
NMR could be used to determine if the spin--spin correlation length
decreases as the semiconducting gap increases.\cite{slichter}
However, our
calculation has neglected certain features of those materials such
as large g-factors and interactions between impurities. 
We have also neglected long range Coulomb interactions and the
associated screening effects which, for example, come into
play between an acceptor ion and the hole it contributes
to the valence band. This is a subject for future study.

\section*{Acknowledgments}
This work was initiated by a stimulating discussion with Z. Fisk. 
We thank S. A. Trugman and H. M. Carruzzo for helpful 
discussions, and
T. Saso for bringing ref. 4 to our attention.
This work was supported in part by ONR grant no. N000014-91-J-1502,
the Center for Materials Science at Los Alamos National Laboratory,
by funds provided by the University of California for the conduct
of discretionary research by Los Alamos National Laboratory,
and an allocation of computer time from the University of 
California, Irvine. 

\section*{Appendix}
In this appendix we show that to zeroth order in perturbation
theory in a periodic system, the spin--spin correlation function 
$<0|S_{z}^{c}(R)S_{z}^{f}(0)|0>=-1/4L$ in a half--filled
one dimensional metal with an odd number of sites and an
Anderson impurity at the center. (Thus there is an
even number of electrons.) $|0>$ is the ground state of the
unperturbed Hamiltonian. To construct the ground state, we note
that there is one electron in the f--orbital of the impurity, 
and one
conduction electron on each site. Since there are an odd
number of sites, there are an odd number of conduction spins. If
we think of filling the states in the conduction band with
conduction electrons, one of the states has an unpaired spin. 
In the ground state the unpaired conduction spin can 
form a singlet or
a triplet with the f-spin. These two states are degenerate since
there are no interactions to zeroth order. Since we know that
the ground state has $S=0$ in the presence of interactions, we
will choose the singlet as the ground state, though we would get
the same result if we chose the triplet as the ground state. Thus,
we can write:
\begin{equation}
|0>=\frac{1}{\sqrt{2}}\left[|\uparrow_{f}\downarrow_{c}>-
|\downarrow_{f}\uparrow_{c}>\right]
\end{equation}
where $|\uparrow_{f}\downarrow_{c}>$ denotes an up f--spin and
a down conduction spin. 

The operator for the z--component of the conduction spin on a site
$R$ is 
\begin{equation}
S_{z}^{c}(R)=\frac{1}{2L}\sum_{k_{1},k_{2}}e^{-i(k_{1}-k_{2})R}
\left(c_{k_{1}\uparrow}^{\dagger}c_{k_{2}\uparrow}-
c_{k_{1}\downarrow}^{\dagger}c_{k_{2}\downarrow}\right)
\end{equation}
$k$ is a good quantum number because the system has periodic
boundary conditions.
To zeroth order, the only 
contribution to $<0|S_{z}^{c}(R)S_{z}^{f}(0)|0>$
comes from the $k_{1}=k_{2}=k_{F}$ term of the
sum. One can show that the other terms in the sum cancel out.
Thus, to lowest order,
\begin{equation}
<0|S_{z}^{c}(R)S_{z}^{f}(0)|0>=-\frac{1}{4L} 
\label{eq:sum}
\end{equation}
%Notice that this result is consistent with the 
%sum rule: \cite{scalapino}
%\begin{equation}
%<0|S_{f}^{2}(0)|0>=-\sum_{R}<0|S_{z}^{c}(R)S_{z}^{f}(0)|0>
%\end{equation}
%since it is easy to show that the right and left hand 
%sides are equal to
%1/4. We have checked that our numerical results obey this sum rule.
Even though we have derived eq. (\ref{eq:sum}) for a periodic
lattice, we expect a similar relation to hold for a chain with 
open boundary conditions, i.e., we expect 
$<0|S_{z}^{c}(R)S_{z}^{f}(0)|0>= -b/L$ to lowest order, 
where the constant $b$ is of order unity.

\begin{figure}

\caption{c-spin--f-spin correlation functions for a symmetric
Anderson impurity
in a metal and a semiconducting host ($V=1$) at half--filling. 
The correlations are between
the f-spin of the impurity and the spins of the conduction electrons.
R is the distance from the impurity site. 
($L=25$, $t=1, \,  V_{0}=1,\, U_{0}=8$). 
Solid lines are guides to the eye.
(a) Linear plot. Notice the Friedel oscillations in the
metallic case. (b) Linear-log plot. In the semiconducting case the 
correlation functions die off very quickly due to 
the presence of the gap.}
\label{fig1}
\vspace*{0.9cm}

\caption{c-spin--f-spin correlation functions for a symmetric 
Anderson impurity
in a semiconducting host at half--filling. The 
correlations are between 
the f-spin of the impurity and the spins of the conduction 
electrons. R is the distance from the impurity site.
($L=25$, $N_{el}=50$, $t=1, \, V_{0}=1,\,U_{0}=8$). The 
correlation length increases as $V$ (and the gap) decreases. 
Solid lines are guides to the eye.}
\label{fig2}

\vspace*{0.9cm}
\caption{A symmetric Anderson impurity in a half--filled 
semiconductor doped with one hole:
(a) Hole density versus site, (b) Spin density versus site. 
For large $V_{0}$ both the hole and 
the spin are spread out over the lattice. For small $V_{0}$ they are 
localized near the impurity which is on site $i=0$. 
$L=25$, $N_{el}=49$, $t=1, \, V=1, \, U_{0}=8$.
Solid lines are guides to the eye.}
\label{fig3}

\vspace*{0.9cm}
\caption{Kinetic energy ($-t<c^{\dagger}_{i \sigma} c_{i+1 \sigma} +
c^{\dagger}_{i+1 \sigma} c_{i\sigma}>$) of a bond between
site $i$ and site $i+1$ versus site $i$ for a hole in
a semiconductor with a symmetric Anderson impurity. 
$L=25$, $N_{el}=49$, $t=1, \, V=1, \, U_{0}=8$.
Notice that the bonds connecting
the impurity site have the largest magnitude of
the kinetic energy for the smallest values of $V_{0}$.
Solid lines are guides to the eye.}
\label{ke}

\vspace*{0.9cm}
\caption{The z-component of the total spin in the f-orbitals and
in the conduction orbitals for a symmetric Anderson impurity
in a half--filled semiconductor (solid symbols, $V=1$, $N_{el}=49$)
doped with a hole versus $V_{0}$. 
The circles are for the total 
f-spin ($S_{z}^{f}=\sum_{i}S_{z}^{f}(i)$) and the squares are for the
total conduction spin ($S_{z}^{c}=\sum_{i}S_{z}^{c}(i)$). 
L=25, $t=1$, and $U_{0}=8$. Solid lines are guides to the eye.}
\label{spinsum}

\vspace*{0.9cm}
\caption{A symmetric Anderson impurity in a half--filled metal
doped with one hole:
(a) Hole density versus site, (b) Spin density versus site.
The hole is never localized
at the impurity site ($i=0$) but the spin density is 
localized for small $V_{0}$.  $L=25$, $N_{el}=25$, $t=1,\, U_{0}=8$.
Solid lines are guides to the eye.}
\label{metalhole}

\vspace*{0.9cm}
\caption{Hole density versus site for a symmetric Anderson
impurity in a half--filled semiconductor doped with two holes.
$t=1$, $V=1$, $U_{0}=8$, $L=25$, $N_{el}=48$, and $S_{z}=0$. For 
$V_{0}=0.1$ and 1.0, one hole is localized in the 
vicinity of the impurity
and the other is spread out over the host lattice. For $V_{0}=0.1$ 
and 1.0, the ground state is nearly degenerate;
the data shown are for the triplet state; 
the data for the singlet state are 
identical. For $V_{0}=2.0$
and 10.0, a singlet forms at the impurity site, and the two holes
are spread over the rest of the lattice. For $V_{0}=2.0$ and 10.0,
the ground state is a nondegenerate singlet.}
\label{fig:2holes}

\vspace*{0.9cm}
\caption{The chemical potential $\mu = E(N)-E(N-1)$
versus the electron filling $N$ for $t=1$, $U_{0}=8$, $V=1$, 
and $L=25$.
The impurity is located in the middle of the symmetric Anderson
lattice. The case of no impurity is shown for comparison. Solid
lines are guides to the eye.}
\label{fig:chempot}

\end{figure}

\begin{thebibliography}{99}

\bibitem{Fradkin} D. Withoff and E. Fradkin, Phys. Rev. Lett.
{\bf 64}, 1835 (1990).
\bibitem{symjap} K. Takegahara, Y. Shimuzu, and O. Sakai, 
J. Phys. Soc. Jpn. {\bf 61}, 3443 (1992).
\bibitem{asymjap} K. Takegahara, Y. Shimuzu, N. Goto, and O. Sakai,
Physica B {\bf 186-188}, 381 (1993).
\bibitem{saso} J. Ogura and T. Saso, J. Phys. Soc. Jpn. {\bf 62},
4364 (1993).
\bibitem{Anderson2} F. D. M. Haldane and P. W. Anderson, 
Phys. Rev. B {\bf 13}, 2553 (1976).
\bibitem{Nishino} T. Nishino and K. Ueda, Phys. Rev. B {\bf 47}, 
12451 (1993).
\bibitem{White} S. R. White, Phys. Rev. Lett. {\bf68}, 3487 (1992);
Phys. Rev. B {\bf48},10345 (1993); R.M.\ Noack, S.R.\ White and D.J.\
Scalapino, in {\it Computer Simulations in Condensed Matter Physics
VII}, Eds. D.P. Landau, K.K. Mon, and H.B. Sch\"uttler (Springer 
Verlag, Heidelberg, Berlin, 1994), p.\ 85-98.
\bibitem{Schrieffer} J. R. Schrieffer and P. A. Wolff, Phys. 
Rev. {\bf 149}, 491 (1966).
\bibitem{guerrero} M. Guerrero and C. C. Yu, Phys. Rev. B {\bf 51},
10301 (1995).
\bibitem{metal} M. Guerrero and R. M. Noack, to be published in
Phys. Rev. B.
\bibitem{gap} For an infinite lattice with periodic boundary 
conditions the
semiconducting gap given by eq. (\ref{eq:gap}) is $\Delta=0.83$
for $V=1$, and $\Delta=0.15$ for $V=0.4$.
\bibitem{error} Our truncation error is reduced to $10^{-4}$ 
because we target
the lowest two states. A targetted state is one whose energy and
wavefunction we try to determine accurately.
\bibitem{degen} The ground state would not be degenerate if the
impurity were not in the middle of a lattice with open boundary
conditions because this corresponds to an asymmetric double well
potential.
%\bibitem{scalapino} R. Blankenbecler, J. R. Fulco, W. Gill, and
%D. J. Scalapino, Phys. Rev. Lett. {\bf 58}, 411 (1987).
\bibitem{furdyna} J. K. Furdyna, J. Appl. Phys. {\bf 64}, R29 (1988).
\bibitem{slichter} J. B. Boyce and C. P. Slichter, Phys. Rev. B 
{\bf 13}, 379 (1976).

\end{thebibliography}
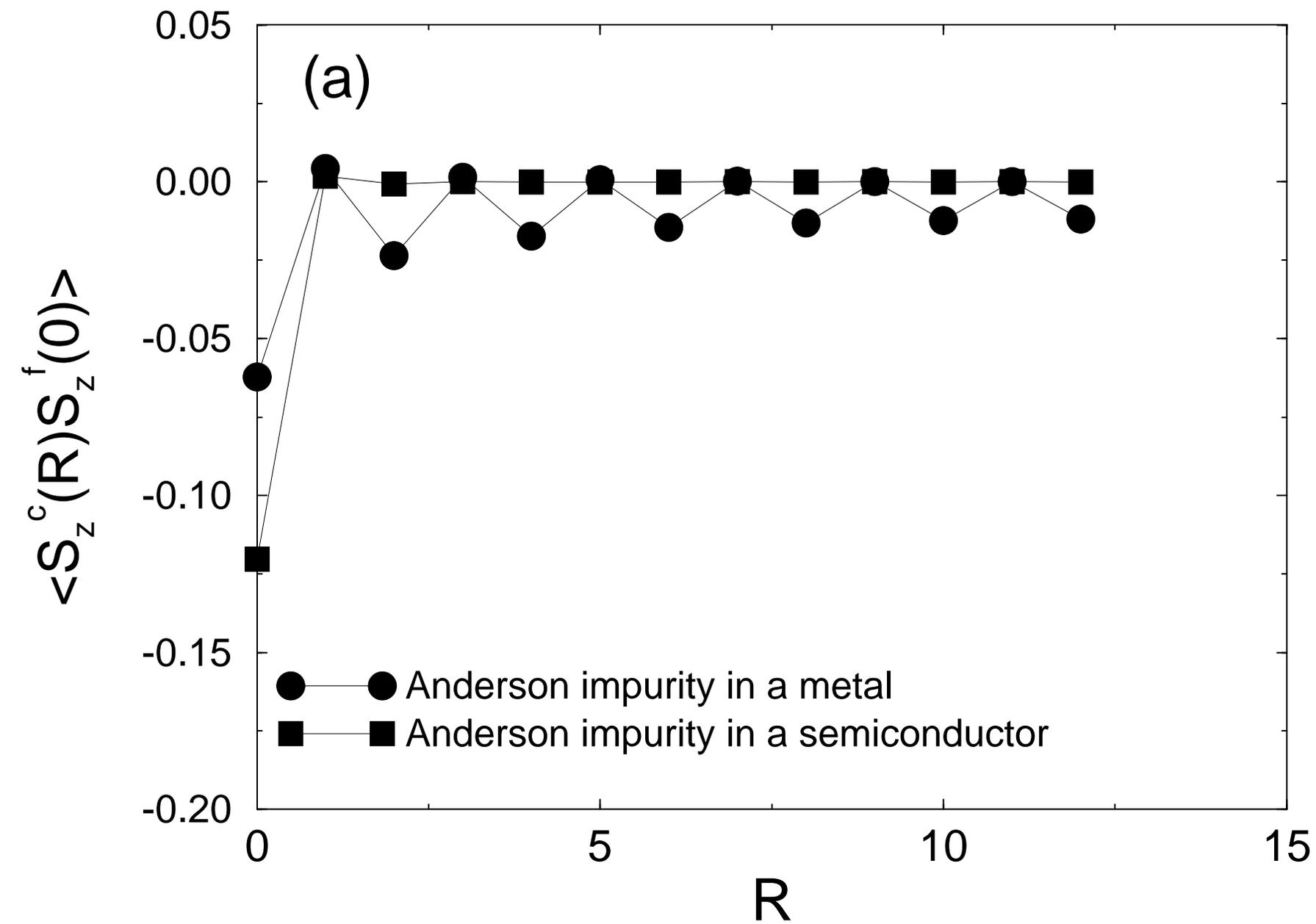
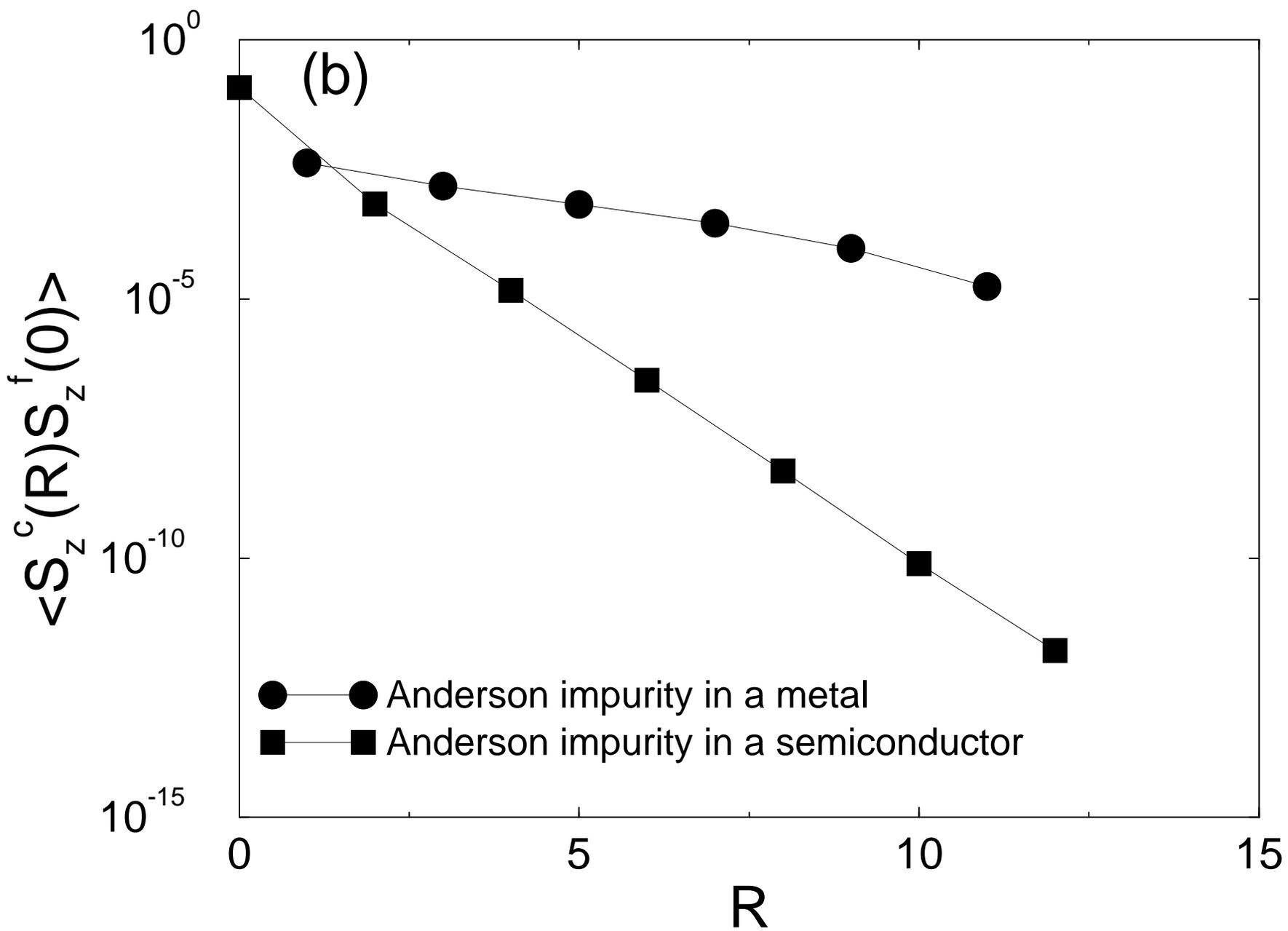
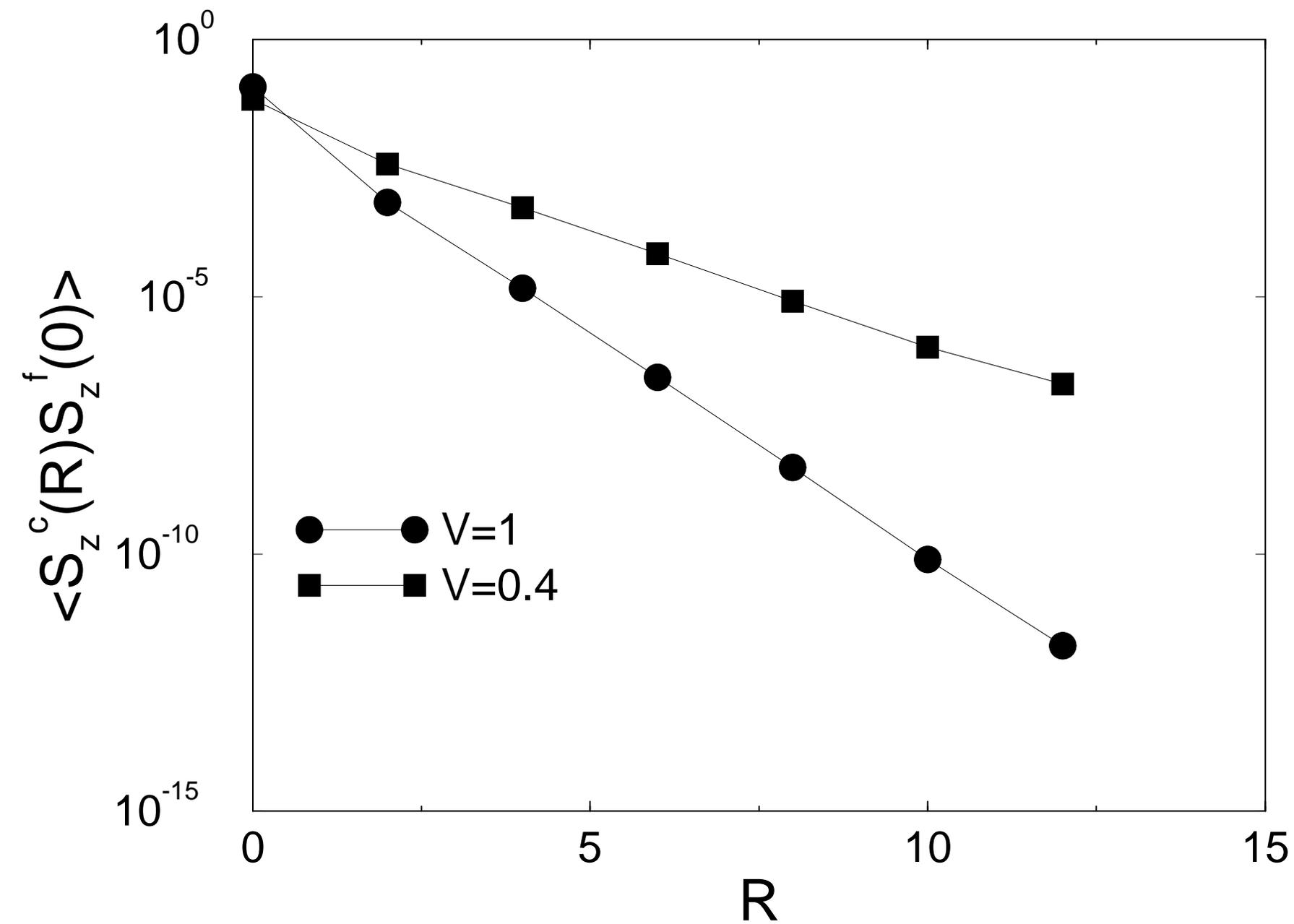
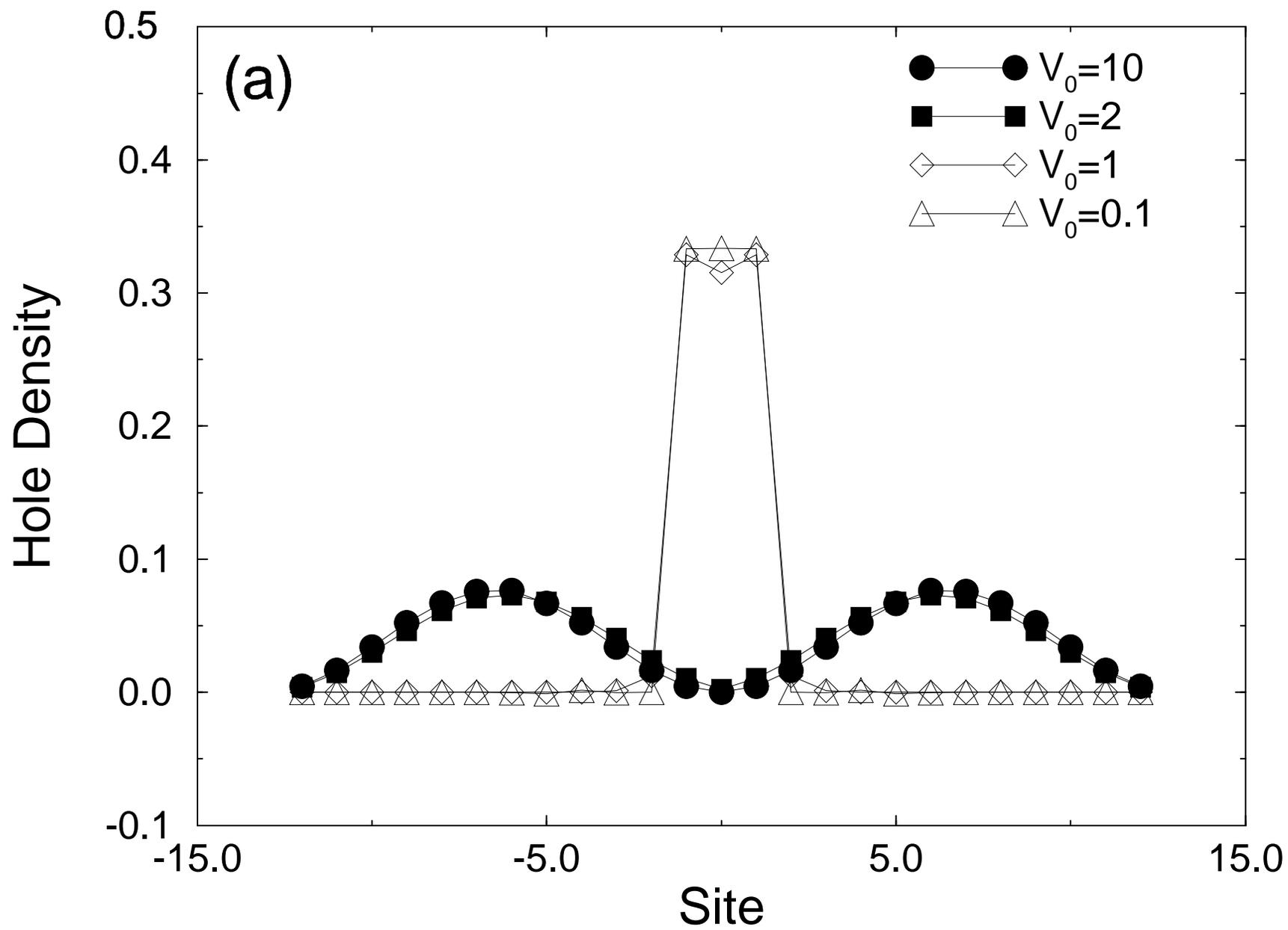
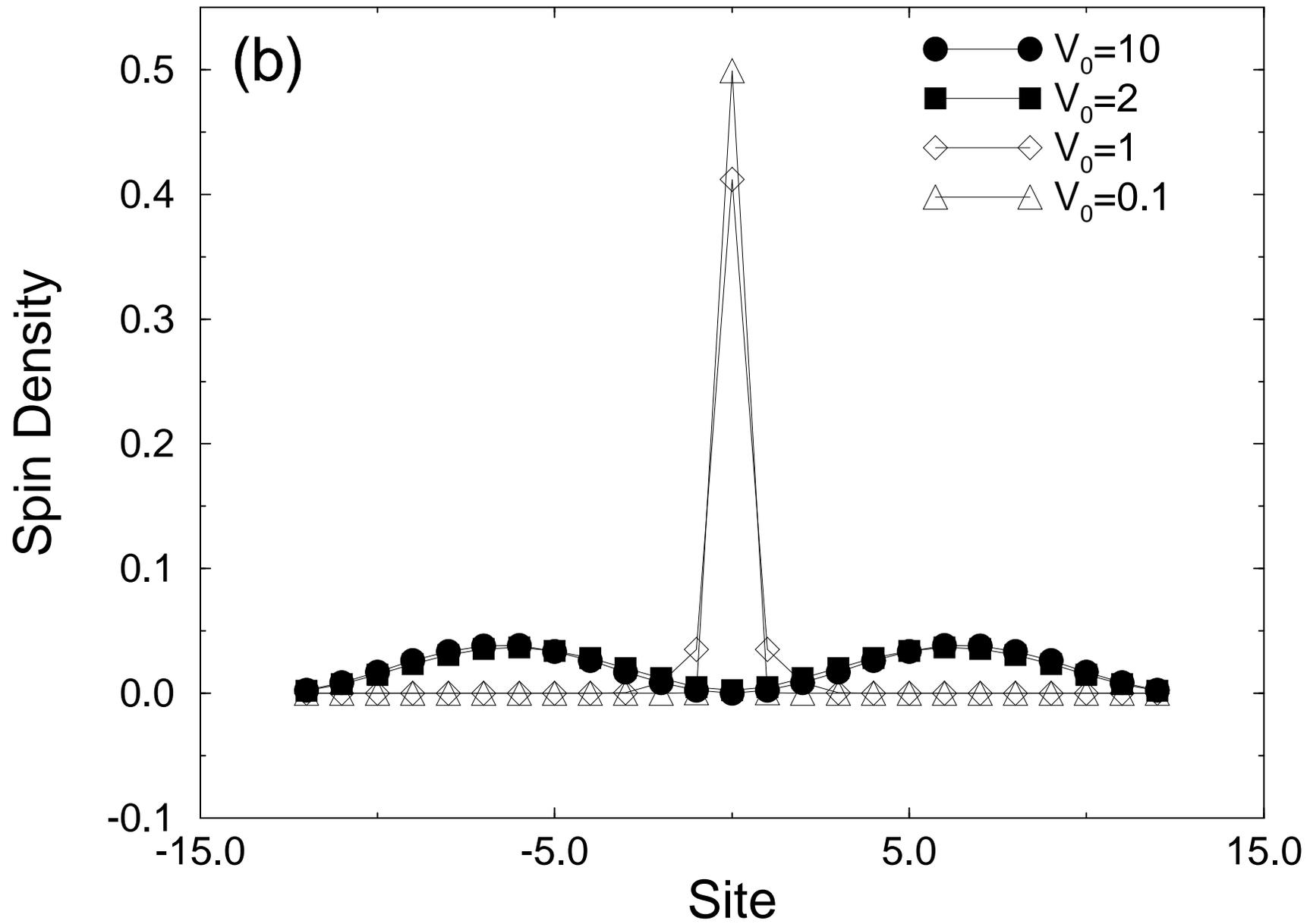
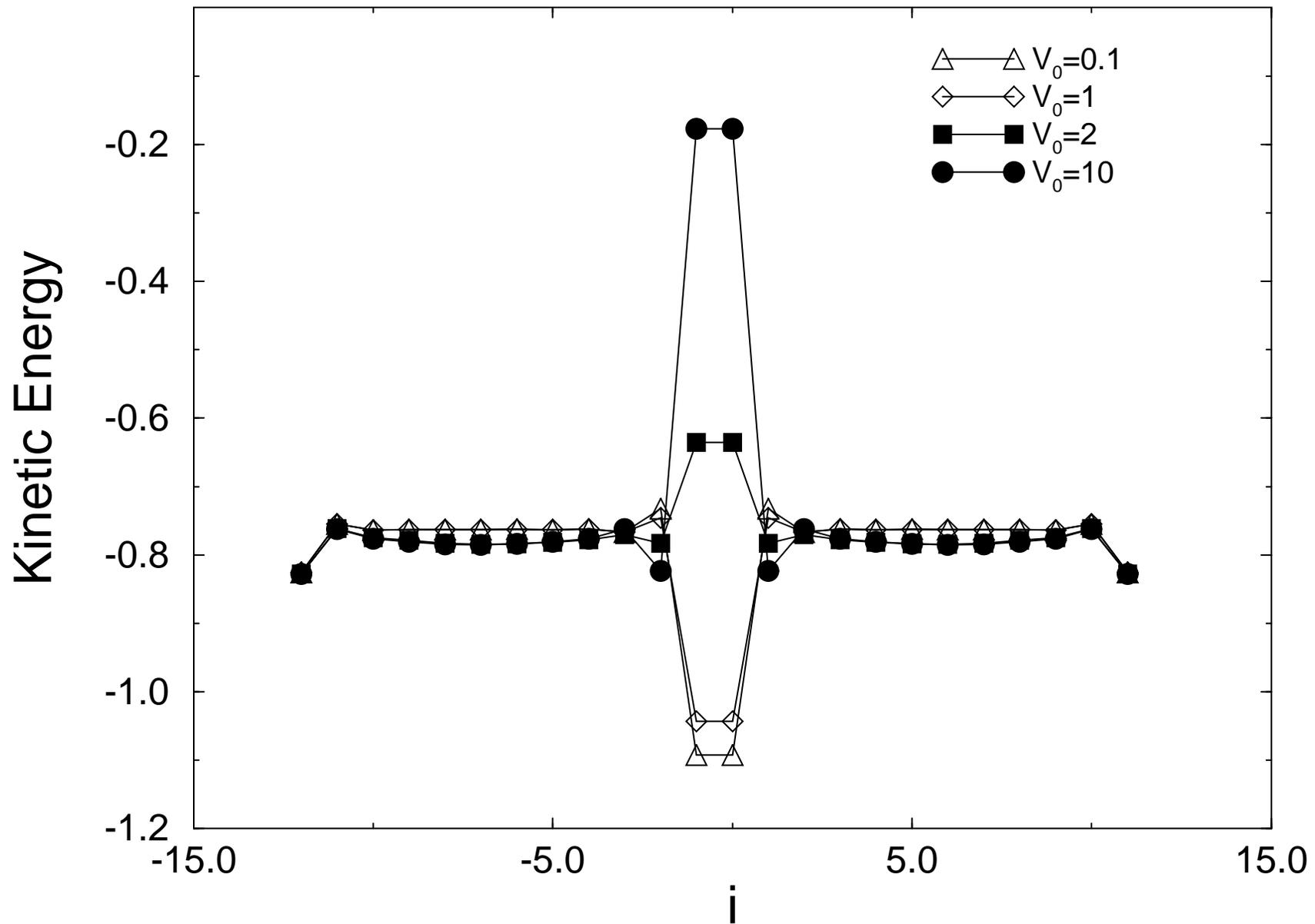
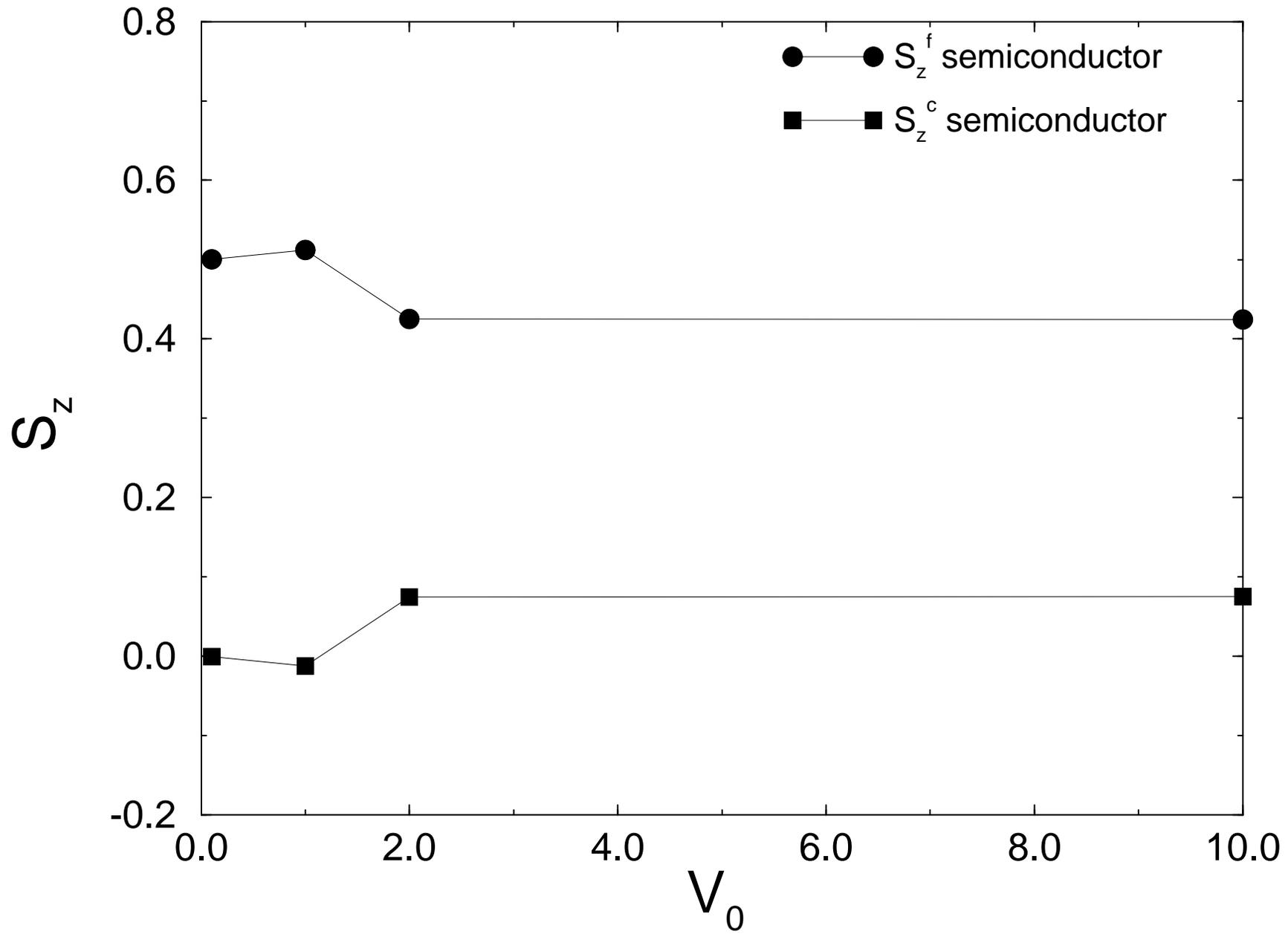
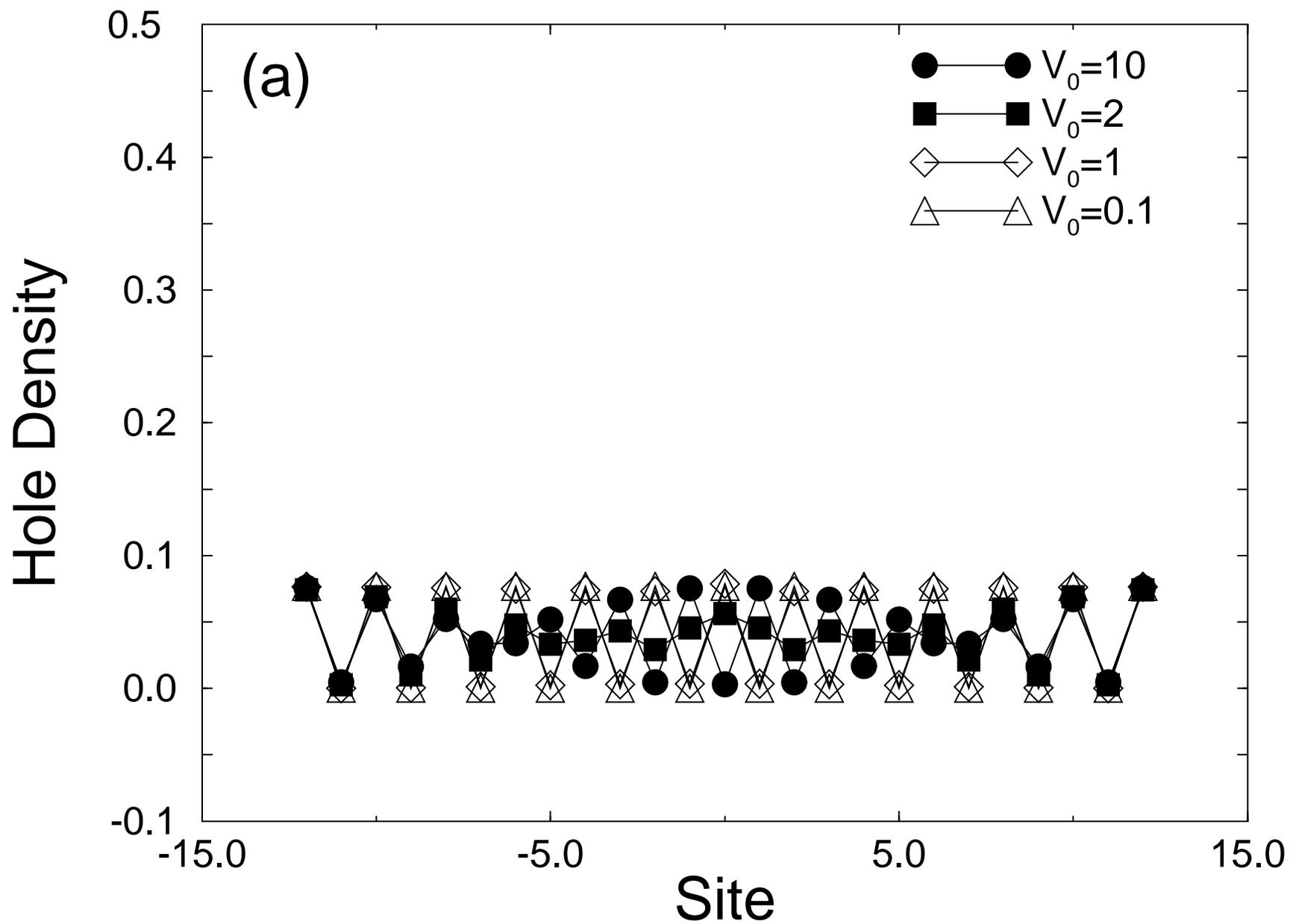
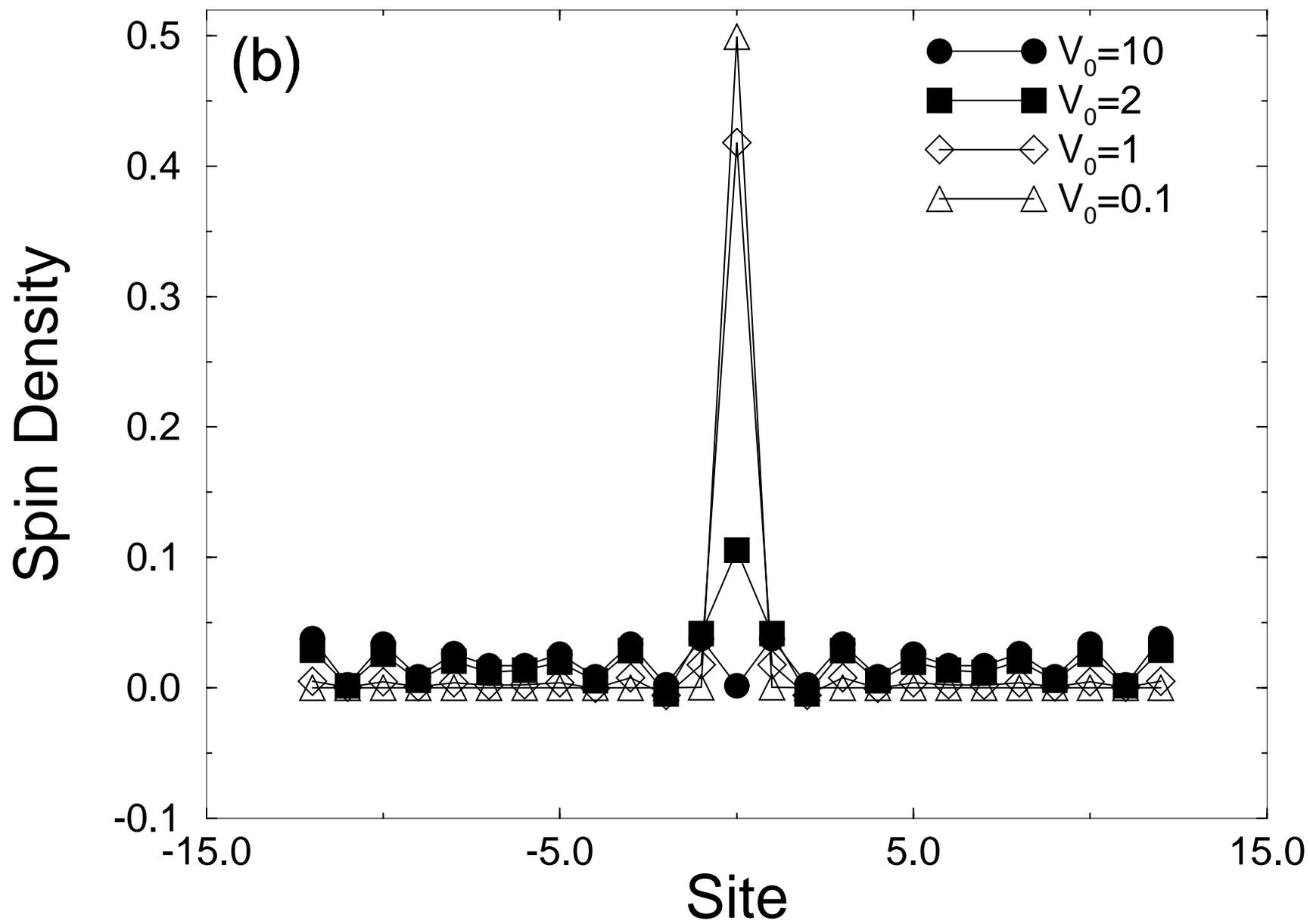
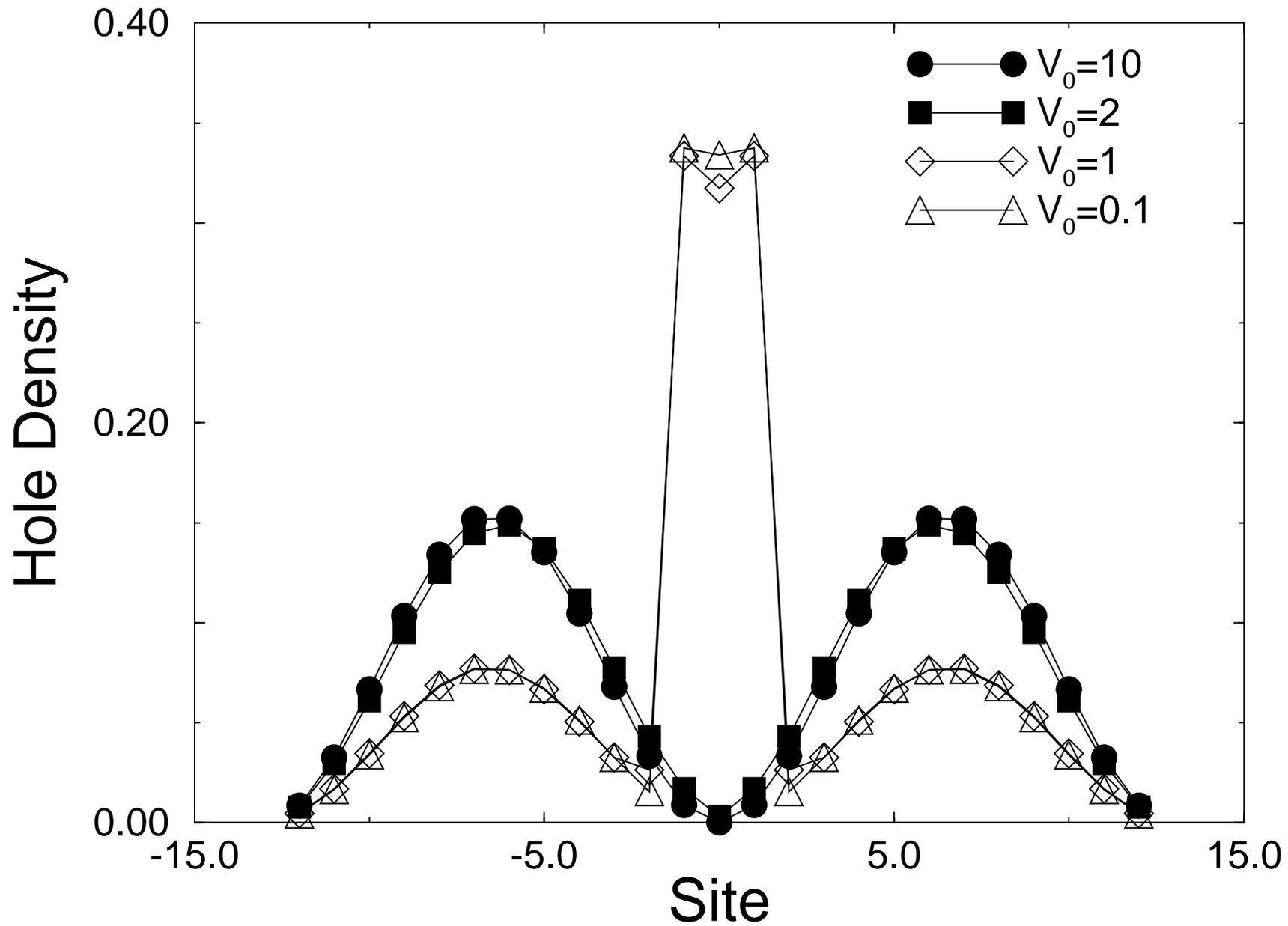
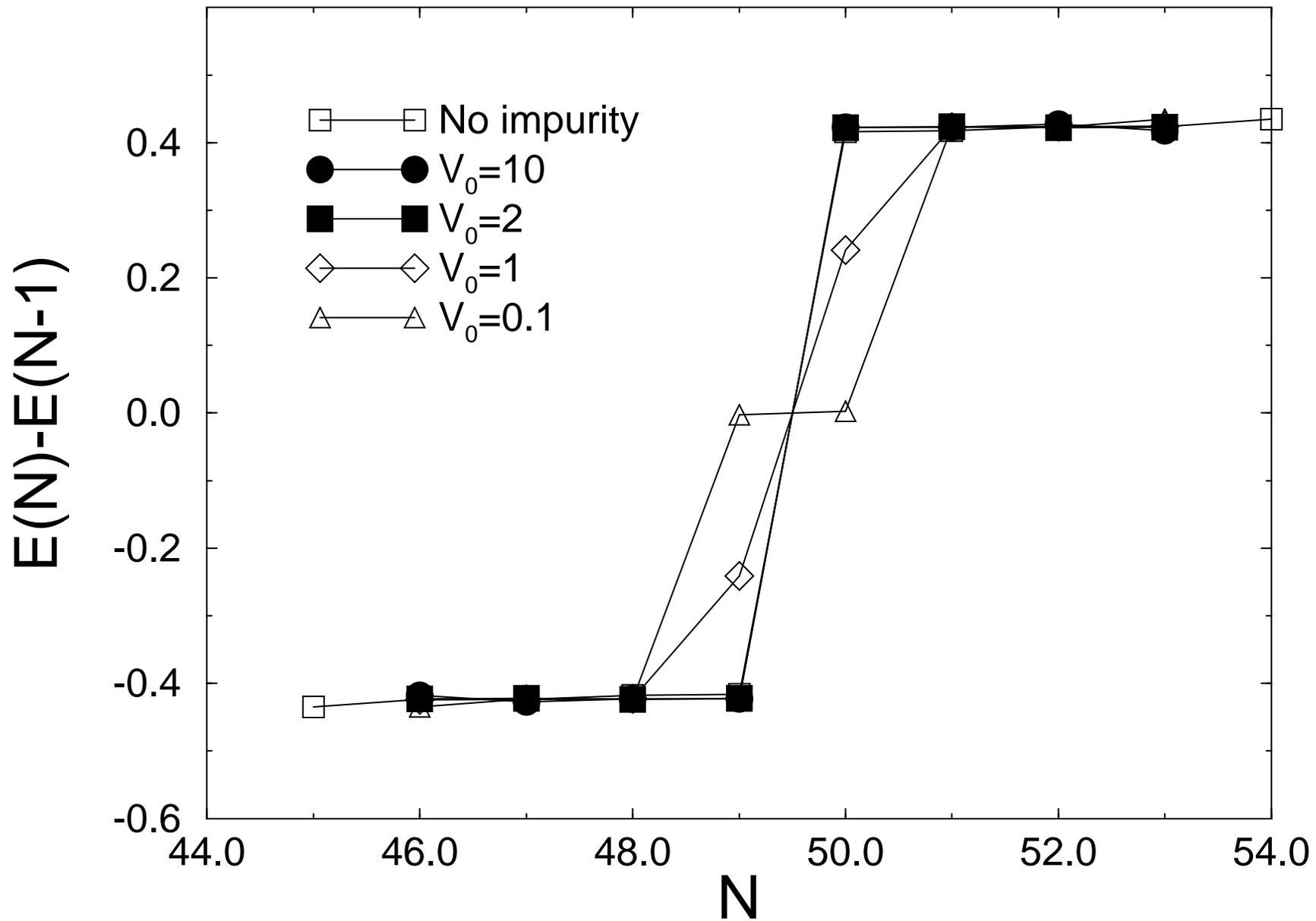
\end{document}